\PassOptionsToPackage{unicode}{hyperref}
\PassOptionsToPackage{hyphens}{url}
\documentclass[
]{article}
\usepackage{xcolor}
\usepackage{amsmath,amssymb}
\usepackage{iftex}
\ifPDFTeX
  \usepackage[T1]{fontenc}
  \usepackage[utf8]{inputenc}
  \usepackage{textcomp} % provide euro and other symbols
\else % if luatex or xetex
  \usepackage{unicode-math} % this also loads fontspec
  \defaultfontfeatures{Scale=MatchLowercase}
  \defaultfontfeatures[\rmfamily]{Ligatures=TeX,Scale=1}
\fi
\usepackage{lmodern}
\ifPDFTeX\else
\fi
\IfFileExists{upquote.sty}{\usepackage{upquote}}{}
\IfFileExists{microtype.sty}{% use microtype if available
  \usepackage[]{microtype}
  \UseMicrotypeSet[protrusion]{basicmath} % disable protrusion for tt fonts
}{}
\makeatletter
\@ifundefined{KOMAClassName}{% if non-KOMA class
  \IfFileExists{parskip.sty}{%
    \usepackage{parskip}
  }{% else
    \setlength{\parindent}{0pt}
    \setlength{\parskip}{6pt plus 2pt minus 1pt}}
}{% if KOMA class
  \KOMAoptions{parskip=half}}
\makeatother
\usepackage{color}
\usepackage{fancyvrb}

\DefineVerbatimEnvironment{Highlighting}{Verbatim}{commandchars=\\\{\}}
\newenvironment{Shaded}{}{}

\newcommand{\DataTypeTok}[1]{\textcolor[rgb]{0.56,0.13,0.00}{#1}}

\newcommand{\ErrorTok}[1]{\textcolor[rgb]{1.00,0.00,0.00}{\textbf{#1}}}

\newcommand{\FunctionTok}[1]{\textcolor[rgb]{0.02,0.16,0.49}{#1}}

\newcommand{\KeywordTok}[1]{\textcolor[rgb]{0.00,0.44,0.13}{\textbf{#1}}}

\newcommand{\OtherTok}[1]{\textcolor[rgb]{0.00,0.44,0.13}{#1}}

\newcommand{\StringTok}[1]{\textcolor[rgb]{0.25,0.44,0.63}{#1}}

\usepackage{longtable,booktabs,array}
\usepackage{calc} % for calculating minipage widths
\usepackage{etoolbox}
\makeatletter
\patchcmd\longtable{\par}{\if@noskipsec\mbox{}\fi\par}{}{}
\makeatother
\IfFileExists{footnotehyper.sty}{\usepackage{footnotehyper}}{\usepackage{footnote}}
\makesavenoteenv{longtable}
\usepackage{graphicx}
\makeatletter
\newsavebox\pandoc@box
\newcommand*\pandocbounded[1]{% scales image to fit in text height/width
  \sbox\pandoc@box{#1}%
  \Gscale@div\@tempa{\textheight}{\dimexpr\ht\pandoc@box+\dp\pandoc@box\relax}%
  \Gscale@div\@tempb{\linewidth}{\wd\pandoc@box}%
  \ifdim\@tempb\p@<\@tempa\p@\let\@tempa\@tempb\fi% select the smaller of both
  \ifdim\@tempa\p@<\p@\scalebox{\@tempa}{\usebox\pandoc@box}%
  \else\usebox{\pandoc@box}%
  \fi%
}
\def\fps@figure{htbp}
\makeatother
\usepackage{graphicx}
\usepackage{float}
\usepackage{xurl}
\usepackage{newunicodechar}
\usepackage{amssymb}
\usepackage{textcomp}

\usepackage{fvextra}
\fvset{breaklines=true, breakanywhere=true, breaksymbol={}}
\RecustomVerbatimEnvironment{verbatim}{Verbatim}{breaklines=true,breakanywhere=true,breaksymbol={}}

\usepackage{etoolbox}
\AtBeginEnvironment{longtable}{\small}

\usepackage{caption}

\usepackage{iftex}

\ifPDFTeX

\newunicodechar{ℝ}{\ensuremath{\mathbb{R}}}

\newunicodechar{⁶}{\textsuperscript{6}}
\newunicodechar{ⁿ}{\textsuperscript{n}}
\newunicodechar{ᵀ}{\textsuperscript{T}}

\newunicodechar{₀}{\textsubscript{0}}
\newunicodechar{₁}{\textsubscript{1}}
\newunicodechar{₂}{\textsubscript{2}}
\newunicodechar{₃}{\textsubscript{3}}
\newunicodechar{₄}{\textsubscript{4}}
\newunicodechar{₅}{\textsubscript{5}}
\newunicodechar{₆}{\textsubscript{6}}
\newunicodechar{₇}{\textsubscript{7}}
\newunicodechar{₈}{\textsubscript{8}}
\newunicodechar{₉}{\textsubscript{9}}

\newunicodechar{ₙ}{\textsubscript{n}}
\newunicodechar{ₖ}{\textsubscript{k}}
\newunicodechar{ₜ}{\textsubscript{t}}
\newunicodechar{ᵢ}{\textsubscript{i}}
\newunicodechar{ⱼ}{\textsubscript{j}}

\newunicodechar{₊}{\textsubscript{+}}
\newunicodechar{₋}{\textsubscript{-}}

\newunicodechar{−}{\ensuremath{-}}
\newunicodechar{∈}{\ensuremath{\in}}
\newunicodechar{∉}{\ensuremath{\notin}}
\newunicodechar{≤}{\ensuremath{\leq}}
\newunicodechar{≥}{\ensuremath{\geq}}
\newunicodechar{≠}{\ensuremath{\neq}}
\newunicodechar{≈}{\ensuremath{\approx}}
\newunicodechar{→}{\ensuremath{\rightarrow}}
\newunicodechar{←}{\ensuremath{\leftarrow}}
\newunicodechar{×}{\ensuremath{\times}}
\newunicodechar{∑}{\ensuremath{\sum}}
\newunicodechar{∞}{\ensuremath{\infty}}

\newunicodechar{ⓘ}{\textcircled{\scriptsize i}}

\newunicodechar{θ}{\ensuremath{\theta}}
\newunicodechar{μ}{\ensuremath{\mu}}
\newunicodechar{σ}{\ensuremath{\sigma}}
\newunicodechar{τ}{\ensuremath{\tau}}
\newunicodechar{φ}{\ensuremath{\varphi}}
\newunicodechar{α}{\ensuremath{\alpha}}
\newunicodechar{β}{\ensuremath{\beta}}
\newunicodechar{ε}{\ensuremath{\varepsilon}}
\newunicodechar{λ}{\ensuremath{\lambda}}
\newunicodechar{γ}{\ensuremath{\gamma}}
\newunicodechar{δ}{\ensuremath{\delta}}
\newunicodechar{ρ}{\ensuremath{\rho}}
\newunicodechar{π}{\ensuremath{\pi}}
\newunicodechar{ω}{\ensuremath{\omega}}

\fi
\usepackage{bookmark}
\IfFileExists{xurl.sty}{\usepackage{xurl}}{} % add URL line breaks if available
\hypersetup{
  hidelinks,
  pdfcreator={LaTeX via pandoc}}

\author{}
\date{}

\begin{document}

\section{Anumati: Proof of Adherence as a Formal Consent Model for
Autonomous Agent
Protocols}\label{anumati-proof-of-adherence-as-a-formal-consent-model-for-autonomous-agent-protocols}

\textbf{Ravi Kiran Kadaboina}

\emph{Independent Researcher}

\emph{M.S., Computer Engineering, University of New Mexico, 2011}

\emph{Anumati} is Sanskrit for ``consent'' or ``formal permission.''

\begin{center}\rule{0.5\linewidth}{0.5pt}\end{center}

\subsection{Abstract}\label{abstract}

As autonomous AI agents increasingly call other agents to complete tasks
on behalf of a human principal, a structural accountability gap has
emerged: the calling agent accepts the terms of service of the callee
without any protocol-level mechanism to prove that it understood those
terms or that it subsequently honoured them. Authentication protocols
such as OAuth and mutual TLS establish \emph{who} may call \emph{which}
capability. They do not address \emph{under what conditions} a permitted
call may be made, and those conditions change as the callee's policies
evolve. In this paper we formalise the distinction between \emph{proof
of acceptance} (a timestamped acknowledgement) and \emph{proof of
adherence} (a per-action reasoning record citing the specific clause
evaluated). We propose three primitives (PolicyDocument, ConsentRecord,
and AdherenceEvent) that together constitute a versioned, append-only
consent model for agent-to-agent communication. The model is
instantiated as a non-breaking extension to two widely used agent
protocols: the Agent2Agent (A2A) protocol and the Model Context Protocol
(MCP). A TLA+ specification of the consent lifecycle, together with a
reference Python implementation of the chain integrity and adherence
trail validators, is available in the accompanying repository.

\begin{center}\rule{0.5\linewidth}{0.5pt}\end{center}

\subsection{1. Introduction}\label{introduction}

In this section we discuss the motivation leading to our work, an
overview of the current state of consent handling in agent protocols,
and our approach to closing the accountability gap that we identify.
Section 1.1 gives background on how calling agents today commit their
principals to the terms of the services they call. Section 1.2 describes
the motivation upon which the work was developed, including the
empirical evidence that agents are not honouring even basic exclusion
signals today, and the regulatory pressure under the EU AI Act. Section
1.3 describes related work in single-agent governance, runtime
enforcement, and privacy standards. Section 1.4 describes the specific
contributions of this work and our approach.

\subsubsection{1.1 Background}\label{background}

The Agent2Agent protocol (A2A) {[}4{]} standardises how one agent
advertises its capabilities and authenticates another. The Model Context
Protocol (MCP) {[}5{]} plays an analogous role for communication between
an orchestrator agent and the tools or data sources it consumes. Both
protocols delegate security to standard HTTP authentication mechanisms
such as OAuth 2.x, API keys, OpenID Connect, and mutual TLS. These
mechanisms are well designed for the question they answer: given a
request, is the requesting agent permitted to invoke the requested
capability.

A second category of rules governs something different. Consider an
agent that provides a data analysis service. Authentication tells the
calling agent whether it is permitted to invoke the
\texttt{analyse\_dataset} skill. It says nothing about whether the
calling agent may store the task output beyond the current session,
whether it may aggregate results across multiple principals, whether it
is obligated to notify its human principal before invoking skills above
a given cost threshold, or whether it may share results with a third
agent the callee has not authorised. These constraints are contextual,
compositional, and they change as the callee's policies evolve. They
cannot be encoded in OAuth scopes, which are discrete, static, and
binary. Scopes answer ``permitted or not.'' Usage policy answers
``permitted, under these conditions, as of this version, subject to
change.''

Further complicating the picture, under the Uniform Electronic
Transactions Act (UETA) §14, contracts formed by electronic agents bind
the human principal, even without their awareness {[}7, 8{]}. As callee
policies evolve and calling agents continue operating under stale terms,
the principal's legal exposure compounds silently.

\subsubsection{1.2 Motivation}\label{motivation}

The gap is not theoretical. The 2025 AI Agent Index surveyed 30 deployed
agents and found that only 6 of them explicitly stated that their
crawler bots respect robots.txt, while 16 provided no clear statement
about web exclusion compliance at all {[}1{]}. Robots.txt is a proxy: if
agents ignore the most basic signal of ``do not access this,'' there is
no reason to expect that they will honour more nuanced usage terms. The
problem is empirical and is worsening as agent deployment scales.

The EU AI Act (Regulation (EU) 2024/1689) compounds the urgency. The Act
enforces in phases. Prohibited AI practices took effect in February 2025
and general-purpose AI model obligations from August 2025. The
obligations most relevant to autonomous agent transactions, namely Art.
14 (human oversight of high-risk AI systems) and Art. 50 (transparency
obligations including disclosure of AI-generated content and automated
decision-making), take effect on August 2, 2026 for high-risk systems
listed in Annex I and Annex III {[}2{]}. As of this writing, the EU AI
Office has published no technical guidance addressing consent mechanisms
for autonomous agent-to-agent transactions {[}3{]}. For systems in
scope, effective oversight requires, at minimum, that the principal can
determine what policies the agent agreed to, what clauses it evaluated
at each action, and what reasoning it applied. The present model is
designed to provide exactly that audit surface in advance of such
guidance, so that implementations are not built retroactively under
deadline pressure.

Several years ago I implemented a versioned terms-of-service system at a
consumer technology company that tried to take this problem seriously.
Policy documents were managed in a headless content management system
(CMS) with full version history, each published version immutable and
content-addressed. The authentication state machine, built on a custom
OAuth 2-based flow, checked on every session whether the user's most
recently accepted document version matched the current published
version. If not, the session was blocked and the user was presented with
the new terms before proceeding. Every acceptance event was stored with
a pointer to the exact document version and a reference to the prior
acceptance record, forming a singly-linked chain per user. The goal was
legal defensibility: in any dispute, the system could show precisely
which version of the terms a user had accepted, and when, across the
full history of their account.

The data was unsurprising. Median time on the acceptance screen was
under four seconds. The linked chain proved acceptance with precision;
it proved nothing about comprehension or subsequent compliance. That
experience motivates the model we propose here. Autonomous agents,
unlike humans, are capable of parsing a policy clause by clause,
evaluating each at every action, and leaving a reasoning trail that a
regulator or principal can inspect after the fact. The consent
infrastructure available to them today does not ask them to do any of
this.

\subsubsection{1.3 Related Work}\label{related-work}

A family of recent proposals addresses agent governance from adjacent
directions. The Open Agent Governance Specification (OAGS) {[}12{]}
defines five governance primitives but is explicitly ``local-first''; it
governs what a single agent may do in its own environment rather than
what two agents agree to when they interact. OpenMandate {[}15{]} takes
a similar single-agent stance using declarative YAML mandates. Policy
Cards {[}16{]} make an agent's constraints inspectable through
machine-readable artefacts, and the \texttt{PolicyDocument} we introduce
is the callee-side equivalent of a Policy Card.

At the runtime-enforcement layer, PCAS {[}17{]} compiles Datalog-derived
policies into instrumented agents that are compliant by construction.
MI9 {[}19{]} detects drift in agent behaviour after the fact. Both
address a different layer than the bilateral consent model we propose; a
deployment could use PCAS for local enforcement and the present model
for cross-agent consent in parallel.

Governance-as-a-Service (GaaS) {[}18{]} proposes an external governance
agent supervising other agents at runtime. The AIGA Internet-Draft
{[}14{]} proposes tiered risk-based governance covering action
authorisation and audit logging. Both govern \emph{what} an agent does;
the model we propose governs \emph{under what agreed terms}.

For financial services, the FINOS AI Governance Framework v2.0 {[}13{]}
defines MI-21 (Agent Decision Audit and Explainability) with tiered
audit logging up to cryptographic tamper-evidence. MI-21 is conceptually
close to the adherence trail we describe in §3.3 but is explicitly
silent on consent versioning.

On the privacy and consent standards side, IEEE P7012 {[}20{]} defines
machine-readable personal privacy terms, and W3C DPV v2.2 {[}21{]}
provides standardised vocabularies for data processing activities. The
Kantara Consent Receipt {[}22{]} established early principles (consent
as receipt, machine-readability, user portability) that inform the
\texttt{ConsentRecord} design in §3.2.

\subsubsection{1.4 Our Approach}\label{our-approach}

We approach the problem by introducing three primitives that together
form a versioned, append-only consent model between any two agents.
\texttt{PolicyDocument} is the callee's machine-readable usage policy,
content-addressed and semver-versioned. \texttt{ConsentRecord} is the
calling agent's parsed understanding of that policy, with an entry for
every clause. \texttt{AdherenceEvent} is a per-action record that cites
the specific clause evaluated, the decision reached, and the
natural-language reasoning the agent applied. The three primitives
together produce two linked lists (a consent chain and an adherence
trail) which together satisfy five properties (completeness,
traceability, tamper evidence, version fidelity, and optional ledger
anchoring) relevant to legal and regulatory accountability.

We instantiate the model as a non-breaking extension to A2A {[}4{]} and
MCP {[}5{]} using each protocol's existing extension mechanism. We
emphasise that the instantiations require no changes to the A2A or MCP
core specifications; both protocols already provide native extension
mechanisms that our protocol, the Agent Consent and Adherence Protocol
(ACAP), uses without modification, and enforcement is carried out by
middleware at the caller and callee boundaries rather than by the core
runtime. We specify the consent lifecycle as a TLA+ state machine and
verify seven safety properties and two liveness properties under the TLC
model checker. A reference Python implementation of the canonical
hashing, chain validator, and adherence trail validator is provided in
the accompanying repository, together with 35 unit tests that exercise
the structural invariants described in §3.

Section 2 discusses why authentication alone is insufficient for usage
policy governance between agents. Section 3 introduces the three
primitives and the formal verification of the consent lifecycle. Section
4 shows how the model integrates into A2A and MCP without modification
of their core specifications. Section 5 surveys related work in more
detail. Section 6 discusses the known limitations of the model,
including the self-attestation boundary and agent ephemerality. Section
7 concludes.

The TLA+ specification, TLC configuration, protobuf schema, and
reference Python implementation are available at:
\url{https://github.com/ravikiran438/agent-consent-protocol}

\begin{center}\rule{0.5\linewidth}{0.5pt}\end{center}

\subsection{2. The Consent Gap in Agent
Protocols}\label{the-consent-gap-in-agent-protocols}

\subsubsection{2.1 What Authentication Covers (and What It Does
Not)}\label{what-authentication-covers-and-what-it-does-not}

A2A supports OAuth 2.x, API keys, OpenID Connect, and mutual TLS
{[}4{]}. MCP uses OAuth 2.1 with incremental scope negotiation following
its November 2025 specification update {[}6{]}. Both are well designed
for their stated purpose: establishing \emph{who} may invoke
\emph{which} capability.

Usage policy governs something different. As described in §1.1, a callee
agent providing a data analysis service cannot express, through OAuth
scopes alone, whether the calling agent may store task output beyond the
current session, aggregate results across multiple principals, notify
its human principal before expensive skill invocations, or share results
with a third agent. These constraints are contextual, compositional, and
evolve as the callee's policies change. They cannot be encoded in OAuth
scopes, which are discrete, static, and binary.

\subsubsection{2.2 The Version Problem}\label{the-version-problem}

Human-facing terms of service change. Services update their policies,
publish new versions, and expect users to re-accept. In practice this is
handled by a login gate: the user sees a banner, clicks agree, and the
session proceeds. The mechanism is crude but functional.

For agents, no equivalent mechanism exists in current protocols. Neither
A2A nor MCP defines a versioned usage policy object, a typed consent
record, or a mechanism to block skill invocation until a new policy has
been processed {[}4, 5{]}. When a callee updates its terms, calling
agents have no protocol-level notification and no structured way to
record re-acceptance. Thus the liability accumulates silently, and the
human principal identified through UETA §14 ends up bound to terms the
agent has never in fact evaluated.

\subsubsection{2.3 Proof of Acceptance vs.~Proof of
Adherence}\label{proof-of-acceptance-vs.-proof-of-adherence}

We distinguish two consent properties.

\textbf{Proof of acceptance} is the traditional model: a timestamped
record binding an identity to a document version at a point in time. It
proves a party \emph{agreed}. It proves nothing about whether the party
\emph{understood} the terms or \emph{subsequently complied} with them.

\textbf{Proof of adherence} is what agents can uniquely provide: a
per-action record citing the specific policy clause evaluated, the
agent's reasoning, and the enforcement decision. It proves the agent
\emph{evaluated} the relevant clause \emph{before acting}.

Table 1 summarises the distinction.

{\def\LTcaptype{none} % do not increment counter
\begin{longtable}[]{@{}lll@{}}
\toprule\noalign{}
Dimension & Proof of Acceptance & Proof of Adherence \\
\midrule\noalign{}
\endhead
\bottomrule\noalign{}
\endlastfoot
Granularity & Whole document & Per-clause, per-action \\
Timing & At acceptance event & At every action attempt \\
Content & Timestamp + identity & Clause citation + reasoning \\
Understanding & Assumed & Verified (parsed claims) \\
Post-acceptance compliance & Unverifiable & Auditable \\
Legal value & Proves agreement & Proves agreement \emph{and}
compliance \\
\end{longtable}
}

Agents are capable of proof of adherence because they can reason about
policy text. The consent infrastructure in current agent protocols
simply does not ask them to. Closing that gap is the work of this paper.

\begin{center}\rule{0.5\linewidth}{0.5pt}\end{center}

\subsection{3. A Formal Consent Model}\label{a-formal-consent-model}

We define three primitives. Together they constitute a versioned,
append-only consent model for agent-to-agent communication. We refer to
the resulting protocol as the \textbf{Agent Consent and Adherence
Protocol (ACAP)}.

\subsubsection{3.1 PolicyDocument}\label{policydocument}

A \texttt{PolicyDocument} is the machine-readable equivalent of a
terms-of-service document. It is versioned using semantic versioning,
content-addressed via SHA-256, and published by the callee agent at a
well-known HTTPS URL.

Formally, a \texttt{PolicyDocument} \emph{P} is a tuple:

\begin{verbatim}
P = (version, hash, effective_date, supersedes, claims, publisher, natural_language_uri)
\end{verbatim}

where \texttt{version} is a semver string, \texttt{hash} is the SHA-256
digest of the canonical JSON serialisation of \emph{P} with the
\texttt{hash} field set to the empty string (which breaks the
circularity such that the digest input never includes itself),
\texttt{effective\_date} is an ISO 8601 timestamp, \texttt{supersedes}
is the version of the document \emph{P} replaces (if any),
\texttt{claims} is a sequence of \texttt{PolicyClaim} objects (order
preserving; claim order is significant for display and diff
computation), \texttt{publisher} is the callee agent's identifier (DID
or HTTPS URL), and \texttt{natural\_language\_uri} is the URL of the
human-readable document from which the claims are derived.

A \texttt{PolicyClaim} \emph{c} is a tuple:

\begin{verbatim}
c = (id, clause_ref, action, asset, rule_type, constraint, since_version,
     category, dimension)
\end{verbatim}

where \texttt{rule\_type} is one of \{permission, prohibition,
obligation\}, \texttt{action} and \texttt{asset} use ODRL 2.2 vocabulary
where applicable {[}9{]}, and \texttt{constraint} is an optional ODRL
constraint expression. The \texttt{since\_version} field records which
policy version introduced this claim, enabling calling agents to compute
diffs between versions. The \texttt{category} and \texttt{dimension}
fields classify the claim by kind of data and kind of operation; these
are consumed by extensions (see §6.3) and may be left unspecified in the
core.

The key design constraint is that every \texttt{PolicyClaim} must have a
stable \texttt{id} across versions. A claim that changes meaning between
versions MUST be assigned a new \texttt{id}; the old \texttt{id} is
retired. This allows calling agents to detect precisely which claims
changed when a version bumps.

\subsubsection{3.2 ConsentRecord}\label{consentrecord}

A \texttt{ConsentRecord} documents the calling agent's parsed
understanding of and decision about a specific \texttt{PolicyDocument}
version. Records form a singly-linked list (the \emph{consent chain})
for a given caller-callee pair.

Formally, a \texttt{ConsentRecord} \emph{R} is a tuple:

\begin{verbatim}
R = (id, prev_id, caller, callee, policy_version, policy_hash,
     parsed_claims, decision, timestamp, valid_until, signature,
     caller_capability_hash, reconsent_trigger)
\end{verbatim}

where \texttt{prev\_id} references the immediately preceding record in
the chain (null for the first record), \texttt{parsed\_claims} contains
one \texttt{ParsedClaim} entry for every \texttt{PolicyClaim} in the
referenced \texttt{PolicyDocument}, \texttt{decision} is one of
\{accepted, rejected, conditional\}, and \texttt{valid\_until} is either
an ISO 8601 timestamp or a sentinel value (\texttt{"on\_version\_bump"},
\texttt{"on\_capability\_change"}, or \texttt{"on\_any\_change"}; see
§3.6).

A \texttt{ParsedClaim} entry records, for each policy claim, whether the
calling agent understood the claim, whether it disputes the claim, and
if disputed, a natural-language explanation of the dispute.

\textbf{Critical invariant}: every \texttt{PolicyClaim} in the
\texttt{PolicyDocument} MUST have a corresponding \texttt{ParsedClaim}
in the \texttt{ConsentRecord}. Agents cannot silently ignore
inconvenient clauses. This single requirement is what separates the
model from a boolean acceptance flag.

Conditional consent is supported such that when \texttt{decision} is
\texttt{conditional}, the calling agent has accepted \emph{some} claims
but disputes others. The callee MUST enforce claim-level gating: a skill
whose governing \texttt{PolicyClaim} is marked \texttt{disputed:\ true}
in the \texttt{ParsedClaim} array is blocked, while skills governed by
undisputed claims remain invocable. This allows the agent to continue
operating at reduced capability rather than halting entirely, which is a
practical necessity for long-running agent workflows where a blanket
rejection would cascade into downstream failures.

The callee enforces claim-level gating by cross-referencing the
\texttt{ParsedClaim} array in the caller's \texttt{ConsentRecord}
against its own skill registry. Each skill in the callee's A2A
\texttt{AgentCard} (or MCP tool manifest) is annotated with the
\texttt{claim\_id} values that govern it via the \texttt{policy\_claims}
array (see §4.1). When the callee receives a task request, it resolves
the caller's active \texttt{ConsentRecord}, iterates the requested
skill's \texttt{policy\_claims}, and checks whether any referenced claim
is marked \texttt{disputed:\ true} or \texttt{understood:\ false}. If
so, the callee MUST reject the invocation with a structured error
indicating which claims block the skill.

The linked-list structure of the consent chain is central to the legal
value of the model. A chain of records shows the full history of what a
calling agent agreed to on behalf of its principal, at the time of each
action, rather than just at signup. We used the same linked structure in
a prior production human-authentication system to preserve all accepted
versions of terms per user for legal auditability. Here it is applied to
the agent-to-agent context and extended to include the per-action
adherence records described in §3.3.

\subsubsection{3.3 AdherenceEvent}\label{adherenceevent}

An \texttt{AdherenceEvent} records the calling agent's runtime
evaluation of a policy claim for a specific action attempt. Events form
a second singly-linked list (the \emph{adherence trail}) anchored to a
\texttt{ConsentRecord}.

Formally, an \texttt{AdherenceEvent} \emph{E} is a tuple:

\begin{verbatim}
E = (id, prev_id, consent_record_id, action, claim_id, clause_ref,
     decision, reasoning, timestamp, context, signature)
\end{verbatim}

where \texttt{decision} is one of \{permit, deny, escalate\} and
\texttt{reasoning} is a natural-language string explaining why the agent
reached that decision.

The \texttt{reasoning} field is the mechanism that makes adherence
auditable. An agent that records:

\begin{quote}
\emph{``Action `aggregate\_sessions' maps to odrl:aggregate on
pii:session\_data. Policy v2.1.0 §3.4 prohibits this where purpose =
behavioural\_profiling. Denying.''}
\end{quote}

has produced something qualitatively different from a system that simply
refuses a request. The reasoning is citable, inspectable, and
attributable. Auditors can trace any enforcement decision back to the
exact clause that governed it.

\subsubsection{3.4 Chain Properties}\label{chain-properties}

The two linked lists (the consent chain and the adherence trail)
together satisfy five properties that are relevant to legal and
regulatory accountability.

\begin{enumerate}
\def\labelenumi{\arabic{enumi}.}
\item
  \textbf{Completeness.} Every action attempt produces an adherence
  event. A deny without an event is indistinguishable from a silent
  failure.
\item
  \textbf{Traceability.} Every adherence event references a consent
  record, which in turn references a policy document version. The full
  chain from action to clause is traversable.
\item
  \textbf{Tamper evidence.} Records and events carry optional JWS
  signatures {[}10{]} over their canonical JSON with the
  \texttt{signature} field set to the empty string prior to signing,
  using the same bootstrapping convention as
  \texttt{PolicyDocument.hash} in §3.1. The resulting signature is then
  inserted into the record. A verifier can confirm that no record was
  modified after signing.
\item
  \textbf{Version fidelity.} The \texttt{policy\_hash} field in the
  consent record binds the acceptance to the exact document content,
  independent of URI availability. The hash survives link rot.
\item
  \textbf{Ledger anchoring (optional).} JWS signatures prove \emph{who
  signed} but not \emph{when}, and they cannot prevent collusion. If
  both parties agree to rewrite the chain, signatures alone cannot
  detect it. For deployments requiring third-party tamper evidence, each
  record may carry a \texttt{ChainAnchor}: a cryptographic hash anchored
  to an external append-only ledger. The model is ledger-agnostic; the
  anchor may point to a public blockchain (Ethereum, Polygon), a
  permissioned ledger (Hyperledger Fabric), or a transparency log (RFC
  9162). Anchoring is strictly additive and deployments that do not
  require it simply omit the \texttt{chain\_anchor} field.
\end{enumerate}

\subsubsection{3.5 Formal Verification of the Consent
Lifecycle}\label{formal-verification-of-the-consent-lifecycle}

We specify the consent lifecycle as a TLA+ state machine and verify its
safety and liveness properties under model checking. The specification
is included in the reference repository
(\texttt{specification/ConsentLifecycle.tla}) with the corresponding TLC
configuration (\texttt{specification/ConsentLifecycle.cfg}).

The lifecycle has seven states for a given caller--callee pair: Idle,
PolicyFetched, GovernanceReview, Accepted, Rejected, Conditional, and
Stale, shown in Figure 1. Table 2 lists the safety and liveness
properties verified by TLC.

\begin{figure}
\centering
\includegraphics[width=0.8\linewidth,height=\textheight,keepaspectratio,alt={ACAP consent lifecycle state machine. The seven states govern the bind/re-bind cycle between a single caller and callee pair. GovernanceReview is reachable only when the governance-tiering extension is loaded (see §6.3); with the core alone, re-consents flow directly from PolicyFetched to the decision states.}]{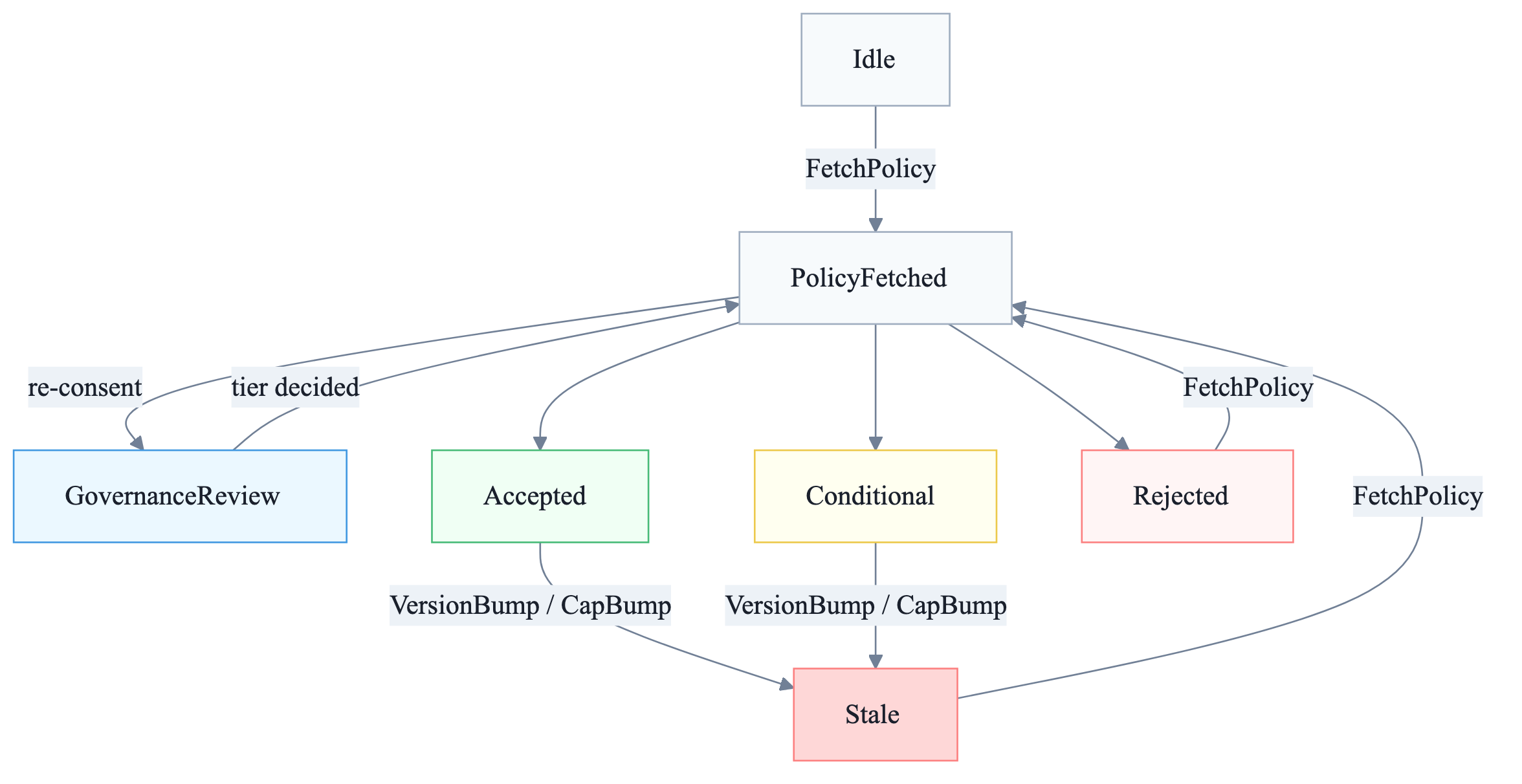}
\caption{ACAP consent lifecycle state machine. The seven states govern
the bind/re-bind cycle between a single caller and callee pair.
GovernanceReview is reachable only when the governance-tiering extension
is loaded (see §6.3); with the core alone, re-consents flow directly
from PolicyFetched to the decision states.}
\end{figure}

\textbf{Table 2.} Safety and liveness properties verified by TLC.

{\def\LTcaptype{none} % do not increment counter
\begin{longtable}[]{@{}
  >{\raggedright\arraybackslash}p{(\linewidth - 6\tabcolsep) * \real{0.1290}}
  >{\raggedright\arraybackslash}p{(\linewidth - 6\tabcolsep) * \real{0.3226}}
  >{\raggedright\arraybackslash}p{(\linewidth - 6\tabcolsep) * \real{0.1935}}
  >{\raggedright\arraybackslash}p{(\linewidth - 6\tabcolsep) * \real{0.3548}}@{}}
\toprule\noalign{}
\begin{minipage}[b]{\linewidth}\raggedright
ID
\end{minipage} & \begin{minipage}[b]{\linewidth}\raggedright
Property
\end{minipage} & \begin{minipage}[b]{\linewidth}\raggedright
Kind
\end{minipage} & \begin{minipage}[b]{\linewidth}\raggedright
Statement
\end{minipage} \\
\midrule\noalign{}
\endhead
\bottomrule\noalign{}
\endlastfoot
S1 & NoSkillWithoutConsent & Safety & A skill call never occurs unless
at least one ConsentRecord in the chain has decision ∈ \{accepted,
conditional\}. \\
S2 & ChainMonotonicity & Safety & \texttt{consentChain} and
\texttt{adherenceTrail} are append-only; lengths never decrease. \\
S3 & AdherenceAnchored & Safety & Every AdherenceEvent references a
valid ConsentRecord index. \\
S4 & SkillRequiresPermit & Safety & A skill call is always preceded by
at least one AdherenceEvent with decision = permit on an undisputed
claim. \\
S5 & ConditionalGating & Safety & A disputed claim always produces deny
or escalate, never permit. \\
S6 & NoDisputedPermit & Safety & No adherence event for a disputed claim
carries a permit decision. \\
S7 & NoSkillOnCapabilityDrift & Safety & No skill call occurs when the
caller's capability fingerprint has changed since the ConsentRecord was
created. \\
L1 & EventualReConsent & Liveness & Under weak fairness, any staleness
(version bump or capability change) eventually leads to a new
ConsentRecord. \\
L2 & EventualCapReConsent & Liveness & A capability bump eventually
leads to re-consent. The agent does not operate indefinitely under stale
reasoning. \\
\end{longtable}
}

Properties S1--S4 establish the core consent-before-action guarantee.
S5--S6 formalise conditional gating for reduced-permission operation. S7
addresses agent ephemerality such that capability drift blocks all skill
invocation until re-consent is obtained. We note that
\texttt{caller\_capability\_hash} is self-reported by the caller and the
callee cannot independently verify it; this is a known limitation which
we discuss in §6.1. L1 and L2 ensure the protocol does not deadlock
after either a policy version bump or a capability change.

The model was checked with TLC using \texttt{MaxVersions\ =\ 3},
\texttt{MaxAdherenceEvents\ =\ 4}, and \texttt{MaxCapVersions\ =\ 2},
covering three policy version bumps, two capability changes, and four
adherence events per consent epoch. All nine properties hold with zero
violations across the reachable state space. TLC performs bounded model
checking such that it exhausts the state space within the declared
constant bounds but does not constitute a proof for arbitrary parameter
values. The bounds were chosen to cover realistic deployment scenarios;
unbounded verification would require inductive proof techniques beyond
the scope of the present work.

\subsubsection{3.6 Capability-Bound
Consent}\label{capability-bound-consent}

Human users persist. They create accounts, accumulate consent history,
and the consent chain tracks a stable identity across years. Agents are
different in kind such that they are spawned, updated, and terminated.
An agent instance may exist for minutes or hours, and the next instance,
even if it shares the same \texttt{caller\_agent\_id}, may run a
different model, carry different tools, or operate under a different
context window.

This creates a problem that human consent systems never face: the
reasoning entity that consented may not be the reasoning entity that
acts. An agent running GPT-4o that recorded \texttt{understood:\ true}
for a prohibition on data aggregation may be replaced by an instance
running a fine-tuned variant that interprets ``aggregation''
differently. The \texttt{ParsedClaim} entries in the original
\texttt{ConsentRecord} are no longer trustworthy, not because they were
falsified, but because the entity that produced them no longer exists.

The model addresses this with \texttt{caller\_capability\_hash}, a
SHA-256 fingerprint of the agent's model identifier, tool manifest, and
reasoning configuration (system prompt, temperature, and any
chain-of-thought or retrieval-augmented generation settings that
influence how the agent interprets policy claims), recorded on every
\texttt{ConsentRecord}. The hash is computed over the canonical JSON
serialisation (RFC 8785) of a \texttt{CapabilityManifest} object
containing these fields in alphabetical key order, using the same
circularity-breaking convention as \texttt{PolicyDocument.hash} in §3.1,
such that the \texttt{caller\_capability\_hash} field is set to the
empty string before hashing. Three re-consent triggers are defined:

{\def\LTcaptype{none} % do not increment counter
\begin{longtable}[]{@{}
  >{\raggedright\arraybackslash}p{(\linewidth - 4\tabcolsep) * \real{0.3750}}
  >{\raggedright\arraybackslash}p{(\linewidth - 4\tabcolsep) * \real{0.2917}}
  >{\raggedright\arraybackslash}p{(\linewidth - 4\tabcolsep) * \real{0.3333}}@{}}
\toprule\noalign{}
\begin{minipage}[b]{\linewidth}\raggedright
Trigger
\end{minipage} & \begin{minipage}[b]{\linewidth}\raggedright
Cause
\end{minipage} & \begin{minipage}[b]{\linewidth}\raggedright
Effect
\end{minipage} \\
\midrule\noalign{}
\endhead
\bottomrule\noalign{}
\endlastfoot
\texttt{POLICY\_BUMP} & Callee publishes new PolicyDocument & Re-fetch,
diff, re-consent \\
\texttt{CAPABILITY\_CHANGE} & Caller's capability hash changes &
Re-evaluate all ParsedClaims \\
\texttt{PRINCIPAL\_CHANGE} & Human principal identity changes & Full
re-consent with new principal\_id \\
\end{longtable}
}

The \texttt{valid\_until} field supports sentinel values that control
invalidation granularity: \texttt{"on\_version\_bump"} for callee-side
changes only, \texttt{"on\_capability\_change"} for caller-side changes
only, or \texttt{"on\_any\_change"} for either side. This lets callees
choose their risk tolerance such that a financial-services agent may
require \texttt{"on\_any\_change"} while a low-stakes utility agent may
accept \texttt{"on\_version\_bump"}.

The consent chain is stored by the callee. It is the callee's audit
trail and the callee's legal protection. The chain is keyed by
\texttt{caller\_agent\_id}, which is a DID or HTTPS URL bound to the
principal or organisation rather than to the process, so new agent
instances inherit the chain if they share the identity. Both parties
SHOULD maintain independent copies of the chain. The callee's copy is
authoritative; the caller's copy provides independent evidence if the
callee's ledger is unavailable, tampered with, or disputed. The chain
itself is append-only; agent termination does not delete records.

\begin{center}\rule{0.5\linewidth}{0.5pt}\end{center}

\subsection{4. Protocol Instantiations}\label{protocol-instantiations}

An important design constraint of the present work is that the
instantiations described in this section do not require changes to the
A2A or MCP core specifications. Both protocols already define a native
extension mechanism (\texttt{capabilities.extensions} in A2A and the
\texttt{capabilities} object in MCP), and ACAP is implemented entirely
through those mechanisms and through middleware at the caller and callee
boundaries. The reference implementation described in §4.3 confirms that
a working end-to-end deployment is buildable today with existing A2A and
MCP libraries, which means adoption is a matter of installing a
middleware library at each participating agent rather than advancing a
specification revision.

\subsubsection{4.1 Agent2Agent (A2A)
Protocol}\label{agent2agent-a2a-protocol}

A2A is an open, vendor-neutral protocol for agent-to-agent communication
{[}4{]}. Each agent's AgentCard is a JSON document published at
\texttt{/.well-known/agent-card.json} that advertises the agent's
capabilities, skills, and authentication requirements.

A2A supports protocol extensions via the
\texttt{capabilities.extensions} array {[}4{]}. Each extension is
declared with a \texttt{uri}, a \texttt{description}, and a
\texttt{required} flag. This is the mechanism used by the Agent Payments
Protocol (AP2) to advertise payment capability {[}11{]}, and we use the
same mechanism for ACAP.

\textbf{Step 1: Declare the extension in the AgentCard}

\begin{Shaded}
\begin{Highlighting}[]
\FunctionTok{\{}
  \DataTypeTok{"capabilities"}\FunctionTok{:} \FunctionTok{\{}
    \DataTypeTok{"extensions"}\FunctionTok{:} \OtherTok{[}\FunctionTok{\{}
      \DataTypeTok{"uri"}\FunctionTok{:} \StringTok{"https://github.com/ravikiran438/agent{-}consent{-}protocol/v0.1"}\FunctionTok{,}
      \DataTypeTok{"description"}\FunctionTok{:} \StringTok{"ACAP v0.1: versioned usage policy and consent auditing."}\FunctionTok{,}
      \DataTypeTok{"required"}\FunctionTok{:} \KeywordTok{true}\FunctionTok{,}
      \DataTypeTok{"params"}\FunctionTok{:} \FunctionTok{\{} \DataTypeTok{"minVersion"}\FunctionTok{:} \StringTok{"0.1"}\FunctionTok{,} \DataTypeTok{"maxVersion"}\FunctionTok{:} \StringTok{"0.1"} \FunctionTok{\}}
    \FunctionTok{\}}\OtherTok{]}
  \FunctionTok{\},}
  \DataTypeTok{"usage\_policy"}\FunctionTok{:} \FunctionTok{\{}
    \DataTypeTok{"version"}\FunctionTok{:} \StringTok{"2.1.0"}\FunctionTok{,}
    \DataTypeTok{"document\_uri"}\FunctionTok{:} \StringTok{"https://callee.example.com/.well{-}known/usage{-}policy.json"}\FunctionTok{,}
    \DataTypeTok{"document\_hash"}\FunctionTok{:} \StringTok{"sha256:a3f5c2..."}\FunctionTok{,}
    \DataTypeTok{"effective\_date"}\FunctionTok{:} \StringTok{"2026{-}02{-}27T00:00:00Z"}\FunctionTok{,}
    \DataTypeTok{"acceptance\_required"}\FunctionTok{:} \KeywordTok{true}\FunctionTok{,}
    \DataTypeTok{"acceptance\_endpoint"}\FunctionTok{:} \StringTok{"https://callee.example.com/acap/consent"}\FunctionTok{,}
    \DataTypeTok{"natural\_language\_uri"}\FunctionTok{:} \StringTok{"https://callee.example.com/terms"}
  \FunctionTok{\}}
\FunctionTok{\}}
\end{Highlighting}
\end{Shaded}

The extension URI is versioned. When a future ACAP version introduces
breaking changes (for example, structured reasoning in a later version),
the URI changes accordingly. A calling agent that encounters an
unrecognised ACAP version SHOULD fall back to the highest mutually
supported version or decline the handshake.

\textbf{Step 2: Annotate skills with governing claims}

Each skill in the AgentCard carries a \texttt{policy\_claims} array
listing the claim IDs that govern its use, mirroring the MCP
\texttt{policyClaims} annotation in §4.2. This allows the calling agent
to evaluate only the relevant claims before invoking each skill:

\begin{Shaded}
\begin{Highlighting}[]
\FunctionTok{\{}
  \DataTypeTok{"skills"}\FunctionTok{:} \OtherTok{[}\FunctionTok{\{}
    \DataTypeTok{"name"}\FunctionTok{:} \StringTok{"analyse\_dataset"}\FunctionTok{,}
    \DataTypeTok{"policy\_claims"}\FunctionTok{:} \OtherTok{[}\StringTok{"claim{-}data{-}retention"}\OtherTok{,} \StringTok{"claim{-}aggregation{-}prohibition"}\OtherTok{]}
  \FunctionTok{\}}\OtherTok{]}
\FunctionTok{\}}
\end{Highlighting}
\end{Shaded}

Without this mapping, the callee has no mechanism to enforce claim-level
gating for conditional consent (§3.2), as it would not know which skills
to block when specific claims are disputed.

\textbf{Step 3: Consent handshake before first skill call}

The calling agent fetches the \texttt{PolicyDocument}, verifies its
hash, parses every \texttt{PolicyClaim}, and POSTs a
\texttt{ConsentRecord} to the \texttt{acceptance\_endpoint} before
invoking any skill. The callee verifies that the record covers every
claim and that the \texttt{policy\_hash} matches the current document.
Only then does it permit skill invocation.

\textbf{Step 4: Per-action adherence recording}

Before each skill call, the calling agent evaluates the relevant claims
and POSTs an \texttt{AdherenceEvent} to the callee. The callee appends
the event to the adherence trail for that consent record. The complete
flow is shown in Figure 2.

\begin{figure}
\centering
\includegraphics[width=0.65\linewidth,height=\textheight,keepaspectratio,alt={ACAP consent and adherence sequence.}]{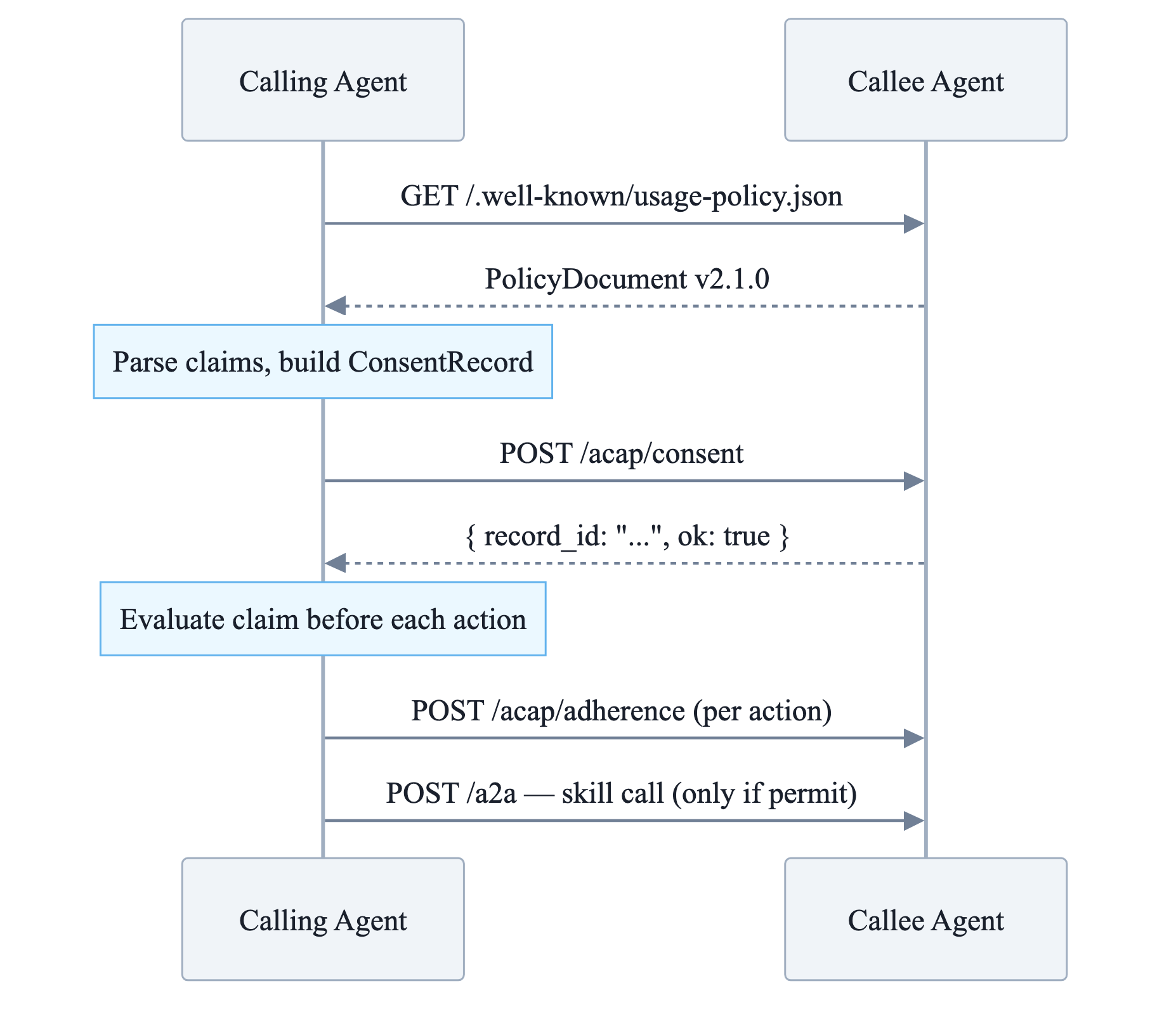}
\caption{ACAP consent and adherence sequence.}
\end{figure}

The \texttt{POST\ /acap/adherence} endpoint operates in one of two
modes, declared in the AgentCard's \texttt{usage\_policy} object. In
\texttt{local} mode the adherence event is fire-and-forget such that the
caller records its own decision and the callee appends it to the audit
trail. In \texttt{delegated} mode the callee evaluates the event and
returns an enforcement decision; the caller MUST NOT invoke the skill
until it receives a \texttt{permit} response. Delegated mode adds a
round trip but gives the callee veto authority, which is appropriate for
regulated interactions where the callee bears compliance obligations.

When the callee publishes a new \texttt{PolicyDocument}, the calling
agent detects the version change on the next AgentCard fetch by
comparing \texttt{usage\_policy.version} against the cached consent
record's \texttt{policy\_version}. It then fetches the new document,
identifies changed claims via \texttt{since\_version}, and creates a new
\texttt{ConsentRecord} with \texttt{prev\_id} pointing to the
now-invalidated record. The old record is never deleted. The current
detection mechanism is poll-based (AgentCard fetch). For long-running
workflows where polling latency is unacceptable, callees SHOULD
additionally support a push channel (webhook callback or SSE endpoint)
registered during the consent handshake.

\subsubsection{4.2 Model Context Protocol
(MCP)}\label{model-context-protocol-mcp}

MCP governs communication between an orchestrator agent (client) and
tools or data sources (servers) {[}5{]}. The MCP specification as of
November 2025 {[}6{]} does not define a consent primitive analogous to
what we propose, and the instantiation below extends MCP using its
standard \texttt{capabilities} object.

The model maps to MCP as follows. The MCP \texttt{initialize} handshake
response includes a \texttt{capabilities} object. We add a
\texttt{usagePolicy} capability:

\begin{Shaded}
\begin{Highlighting}[]
\FunctionTok{\{}
  \DataTypeTok{"capabilities"}\FunctionTok{:} \FunctionTok{\{}
    \DataTypeTok{"tools"}\FunctionTok{:} \FunctionTok{\{\},}
    \DataTypeTok{"usagePolicy"}\FunctionTok{:} \FunctionTok{\{}
      \DataTypeTok{"version"}\FunctionTok{:} \StringTok{"1.0.0"}\FunctionTok{,}
      \DataTypeTok{"documentUri"}\FunctionTok{:} \StringTok{"https://server.example.com/.well{-}known/usage{-}policy.json"}\FunctionTok{,}
      \DataTypeTok{"documentHash"}\FunctionTok{:} \StringTok{"sha256:b7f3a1..."}\FunctionTok{,}
      \DataTypeTok{"acceptanceRequired"}\FunctionTok{:} \KeywordTok{true}
    \FunctionTok{\}}
  \FunctionTok{\}}
\FunctionTok{\}}
\end{Highlighting}
\end{Shaded}

Individual tool definitions can carry a \texttt{policyClaims} annotation
listing the claim IDs that govern their use:

\begin{Shaded}
\begin{Highlighting}[]
\FunctionTok{\{}
  \DataTypeTok{"name"}\FunctionTok{:} \StringTok{"aggregate\_user\_data"}\FunctionTok{,}
  \DataTypeTok{"description"}\FunctionTok{:} \StringTok{"Aggregates user data across sessions."}\FunctionTok{,}
  \DataTypeTok{"policyClaims"}\FunctionTok{:} \OtherTok{[}\StringTok{"claim{-}pii{-}aggregate{-}prohibition"}\OtherTok{]}\FunctionTok{,}
  \DataTypeTok{"inputSchema"}\FunctionTok{:} \FunctionTok{\{} \ErrorTok{...} \FunctionTok{\}}
\FunctionTok{\}}
\end{Highlighting}
\end{Shaded}

This allows calling agents to evaluate only the relevant subset of
claims before invoking each tool rather than re-evaluating the entire
policy document per call. The consent and adherence records use the same
schema as the A2A instantiation.

The MCP instantiation is lighter because MCP's trust boundary is
typically within a single organisation (orchestrator to owned tools).
Cross-organisational A2A calls carry higher legal weight and warrant the
fuller handshake described in §4.1.

\subsubsection{4.3 Reference Implementation and
Overhead}\label{reference-implementation-and-overhead}

The accompanying repository provides a Python reference implementation
of the three structural operations that any ACAP implementation must
perform on every interaction: canonical hashing of the
\texttt{PolicyDocument} per §3.1, structural validation of the consent
chain per §3.2, and structural validation of the adherence trail per
§3.3--§3.4. The implementation is accompanied by 35 unit tests that
exercise the invariants described in those sections.

To characterise the per-call cost of the protocol on commodity hardware,
we ran a micro-benchmark of each operation on a 2024 MacBook Air (Apple
M4, Python 3.13.5). Each measurement is the median of 200 to 500
samples; the 99th percentile is reported alongside. The policy used for
the chain-validation rows contains ten claims, which is a typical size
for a production usage policy.

{\def\LTcaptype{none} % do not increment counter
\begin{longtable}[]{@{}lrr@{}}
\toprule\noalign{}
Operation & Median & p99 \\
\midrule\noalign{}
\endhead
\bottomrule\noalign{}
\endlastfoot
\texttt{compute\_policy\_hash} (10 claims) & 19 μs & 28 μs \\
\texttt{compute\_policy\_hash} (50 claims) & 74 μs & 88 μs \\
\texttt{compute\_policy\_hash} (200 claims) & 278 μs & 324 μs \\
\texttt{validate\_consent\_chain} (chain length 1) & 20 μs & 27 μs \\
\texttt{validate\_consent\_chain} (chain length 5) & 96 μs & 124 μs \\
\texttt{validate\_consent\_chain} (chain length 20) & 373 μs & 458 μs \\
\texttt{validate\_adherence\_trail} (trail length 10) & 3 μs & 9 μs \\
\texttt{validate\_adherence\_trail} (trail length 100) & 27 μs & 32
μs \\
\texttt{validate\_adherence\_trail} (trail length 1000) & 265 μs & 331
μs \\
\end{longtable}
}

These numbers are in microseconds. A typical remote skill invocation
between two agents over the public internet sits in the 20--100 ms range
such that the ACAP overhead measured here represents well under one
percent of the skill-call latency it accompanies, even at chain and
trail lengths that are conservative overestimates of steady-state
operation.

Two caveats apply. First, the benchmark measures only the in-process
structural operations and does not include the network round-trip for
\texttt{POST\ /acap/consent} or \texttt{POST\ /acap/adherence}; the
actual wall-clock cost of the consent handshake on the wire will be
dominated by TCP and TLS setup rather than by the validators themselves.
Second, JWS signature verification is not yet implemented in the
reference validators and is therefore excluded from the numbers above.
We expect signature verification to add a further tens of microseconds
per record on the same hardware; this is a natural next step for the
reference implementation.

\subsubsection{4.4 Demo Deployment}\label{demo-deployment}

To confirm that the middleware framing in §4 actually composes with
existing A2A infrastructure, we built a two-agent end-to-end deployment.
The callee is a data-analysis agent implemented as a FastAPI service
that publishes an A2A AgentCard at \texttt{/.well-known/agent-card.json}
and mounts the ACAP callee middleware described in
\texttt{acap.middleware.callee} at the \texttt{/acap} URL prefix. The
caller is a marketing-insights agent implemented as an async Python
client wrapped in the \texttt{ACAPCaller} described in
\texttt{acap.middleware.caller}. Both agents use Gemini 2.5 Flash as
their reasoning backend such that the callee's skill delegates its
one-shot dataset analysis to Gemini and the caller uses Gemini to parse
each \texttt{PolicyClaim} into a \texttt{ParsedClaim} during the
handshake.

The demonstration scenario is a policy with three prohibitions: a
retention limit on session data, a prohibition against aggregation for
behavioural profiling, and a prohibition against third-party
distribution. The caller declares its intent as producing a marketing
insights report that does not engage in behavioural profiling of
individual customers. Gemini parses each of the three claims as
understood and undisputed under that intent such that the caller arrives
at a \texttt{ConsentRecord} with \texttt{decision\ =\ accepted}. The
caller then attempts two skill calls in sequence: the first with
\texttt{purpose\ =\ behavioural\_profiling}, which matches the
prohibition's constraint and is denied at the adherence layer before the
skill is ever invoked, and the second with
\texttt{purpose\ =\ statistical\_analysis}, which the prohibition
permits and which returns a qualitative summary.

\begin{figure}
\centering
\includegraphics[width=0.95\linewidth,height=\textheight,keepaspectratio,alt={Caller agent trace: consent handshake (top), blocked skill call on a disputed purpose (middle), and permitted skill call returning a qualitative summary (bottom). Each parsed claim is produced by a Gemini call, and each adherence decision records a natural-language reasoning string as described in §3.3.}]{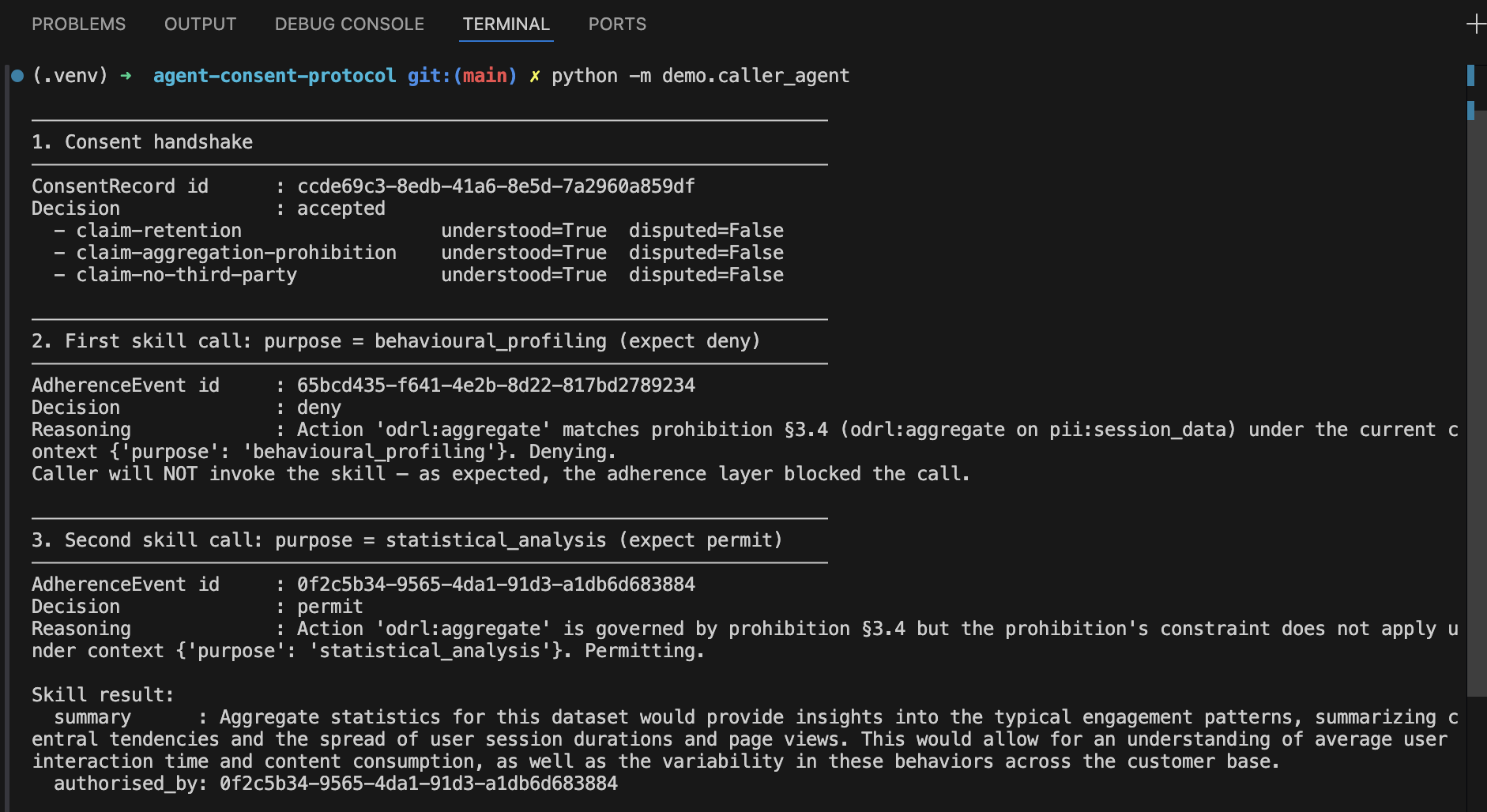}
\caption{Caller agent trace: consent handshake (top), blocked skill call
on a disputed purpose (middle), and permitted skill call returning a
qualitative summary (bottom). Each parsed claim is produced by a Gemini
call, and each adherence decision records a natural-language reasoning
string as described in §3.3.}
\end{figure}

Figure 3 shows the caller's trace across the consent handshake and both
skill calls. The handshake block displays the three parsed claims as
evaluated by Gemini; the deny block shows the reasoning string recorded
by the \texttt{AdherenceEvent}; and the permit block shows the callee's
skill response alongside the \texttt{event\_id} that authorised it.

\begin{figure}
\centering
\includegraphics[width=0.95\linewidth,height=\textheight,keepaspectratio,alt={Callee server log for the same end-to-end session. Every interaction is a standard HTTP request against a FastAPI app, with the ACAP endpoints (/acap/consent, /acap/adherence, /acap/audit) sitting alongside the A2A-standard well-known paths. No modification of the A2A runtime is required.}]{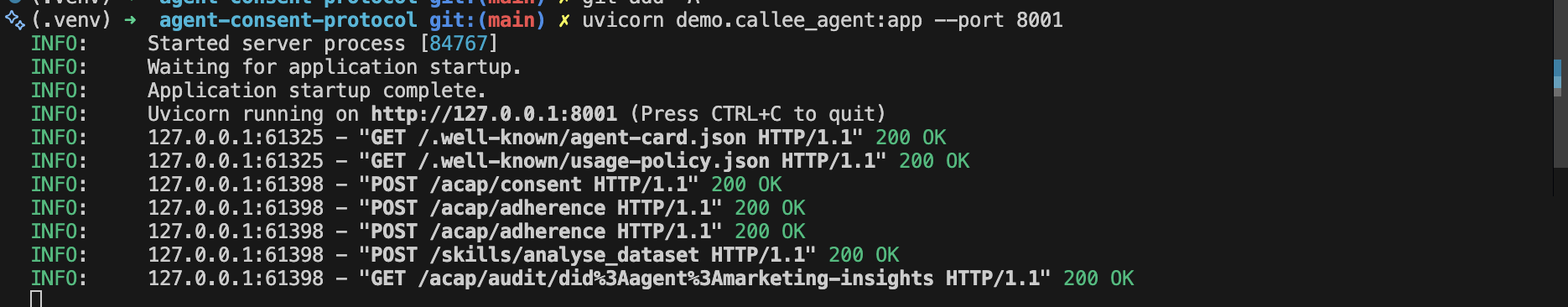}
\caption{Callee server log for the same end-to-end session. Every
interaction is a standard HTTP request against a FastAPI app, with the
ACAP endpoints (\protect\texttt{/acap/consent},
\protect\texttt{/acap/adherence}, \protect\texttt{/acap/audit}) sitting
alongside the A2A-standard well-known paths. No modification of the A2A
runtime is required.}
\end{figure}

Figure 4 shows the callee's HTTP log over the same session, which is the
cleanest evidence that ACAP requires no protocol-level modification:
every interaction is a standard HTTP request, and the ACAP-specific
endpoints sit alongside the A2A-standard
\texttt{/.well-known/agent-card.json} and
\texttt{/.well-known/usage-policy.json} paths under one FastAPI app.

\begin{figure}
\centering
\includegraphics[width=0.85\linewidth,height=\textheight,keepaspectratio,alt={Audit endpoint output. The linked-list structure of the consent chain and adherence trail is visible through the prev\_record\_id and prev\_event\_id fields, and the per-action reasoning is preserved verbatim for later inspection.}]{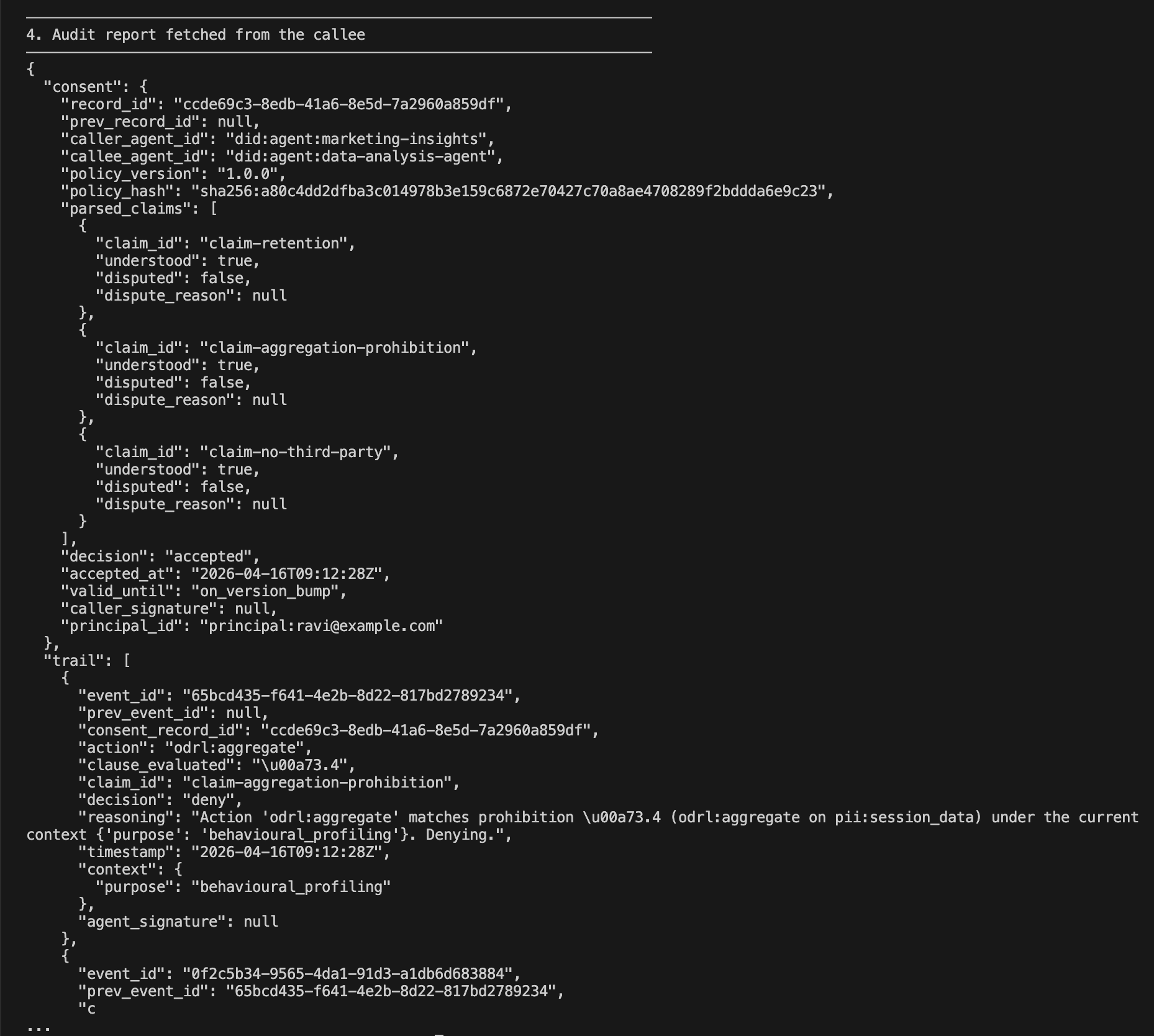}
\caption{Audit endpoint output. The linked-list structure of the consent
chain and adherence trail is visible through the
\protect\texttt{prev\_record\_id} and \protect\texttt{prev\_event\_id}
fields, and the per-action reasoning is preserved verbatim for later
inspection.}
\end{figure}

Figure 5 shows the audit endpoint's response. The linked-list structure
of the consent chain and the adherence trail is visible through the
\texttt{prev\_record\_id} and \texttt{prev\_event\_id} fields, and the
per-action reasoning is preserved verbatim for inspection by a human
principal, a compliance officer, or a regulator.

The full deployment code is in the \texttt{demo/} directory of the
accompanying repository. A reviewer can reproduce Figures 3 through 5 by
following the README there; the end-to-end run completes in under ten
seconds on the same MacBook Air M4 that produced the benchmark numbers
in §4.3. The wall-clock time is dominated by the Gemini inference calls
(three during the handshake, one for the permitted skill invocation)
rather than by the ACAP validators themselves, which contribute the
microsecond-scale overhead reported in §4.3.

The Gemini-backed claim parser used in the demo is illustrative, not
normative. Because the parser is a non-deterministic language model, two
runs of the handshake against the same \texttt{PolicyDocument} can in
principle produce two different \texttt{ParsedClaim} arrays such that
the resulting \texttt{ConsentRecord} values are not byte-identical and
therefore not reproducible across runs. A production deployment should
instead use a deterministic parser, for example a rule-based ODRL
evaluator, a cached LLM output keyed on
\texttt{(claim\_id,\ caller\_intent)} hash, or a constrained-decoding
pipeline with a fixed random seed. The caller middleware is designed to
accept any implementation of the \texttt{ClaimParser} protocol such that
an operator can substitute a deterministic parser without changes to any
other component. The Gemini-backed parser remains useful during agent
development for exploration and for qualitative evaluation of a
candidate policy.

\begin{center}\rule{0.5\linewidth}{0.5pt}\end{center}

\subsection{5. Related Work}\label{related-work-1}

A fuller treatment of related work complements the summary in §1.3. We
survey proposals by the layer each addresses.

\textbf{Single-agent governance.} OAGS {[}12{]} defines five governance
primitives (deterministic identity, declarative policy, runtime
enforcement, structured audit evidence, cryptographic verification) but
is explicitly ``local-first.'' It governs what one agent may do in its
own environment, not what two agents agree to when they interact.
OpenMandate {[}15{]} takes a similar single-agent stance, separating
policy from execution using declarative YAML mandates. Both address the
right abstraction level but stop at the organisational boundary. Policy
Cards {[}16{]} make an individual agent's constraints inspectable
through machine-readable governance artefacts. The
\texttt{PolicyDocument} is the callee's equivalent of a Policy Card, but
\texttt{ConsentRecord} and \texttt{AdherenceEvent} capture the bilateral
negotiation and ongoing compliance that Policy Cards do not cover.

\textbf{Runtime enforcement.} PCAS {[}17{]} compiles Datalog-derived
policies into instrumented agents that are policy-compliant by
construction. A deployment could use PCAS for local enforcement and ACAP
for cross-agent consent; the two address different layers. MI9 {[}19{]}
focuses on drift detection, identifying when agents deviate from
expected behaviour. MI9 detects non-compliance after the fact; ACAP
provides the evidentiary basis (what was agreed, what was checked) that
drift detection consumes.

\textbf{Governance architecture.} GaaS {[}18{]} proposes an external
governance agent supervising other agents at runtime. The AIGA
Internet-Draft {[}14{]} proposes tiered risk-based governance covering
action authorisation and audit logging. AIGA governs \emph{what} an
agent does; ACAP governs \emph{under what agreed terms}.

\textbf{Financial services.} FINOS AI Governance Framework v2.0 {[}13{]}
defines MI-21 (Agent Decision Audit and Explainability) with tiered
audit logging up to cryptographic tamper-evidence. MI-21 is conceptually
close to the adherence trail but is explicitly silent on consent
versioning {[}13{]}. A FINOS contribution to address the versioning gap
is in preparation.

\textbf{Privacy and consent standards.} IEEE P7012 {[}20{]} defines
machine-readable personal privacy terms. P7012 addresses the
human-to-service relationship; ACAP addresses the agent-to-agent
relationship where human preferences must propagate through delegation
chains. W3C DPV v2.2 {[}21{]} provides standardised vocabularies for
data processing activities. The Kantara Consent Receipt {[}22{]}
established early principles (consent as receipt, machine-readability,
user portability) that inform the \texttt{ConsentRecord} design.

\textbf{OAuth ecosystem.} UMA 2.0 {[}26{]} enables resource owners to
control delegated access through an authorisation server. UMA governs
\emph{who may access which resource}; ACAP governs \emph{what a calling
agent agreed the callee may do} and whether it honoured those terms at
every action. RFC 9396 (Rich Authorization Requests) {[}25{]} extends
OAuth with structured authorization data beyond binary scopes. RAR lets
a client express \emph{what} it wants to do, but is a request-time
mechanism with no versioning, no per-action adherence, and no re-consent
on policy change.

\textbf{Audit trails.} AuditableLLM {[}23{]} contributes hash-chain
audit trails for language model interactions. Our chain structure shares
the pattern but applies it to consent and policy adherence across agent
boundaries.

\textbf{Foundational analysis.} Rida {[}24{]} examines what consent
means when an AI agent, not a human, is the party clicking ``I agree.''
That analysis directly motivates the present work.

\textbf{ODRL 2.2} {[}9{]} provides the vocabulary we use for
\texttt{PolicyClaim.action} and \texttt{PolicyClaim.asset}. We treat
ODRL as a vocabulary convention, not a normative dependency. ODRL is a
rights \emph{expression} language, not a consent \emph{lifecycle}
protocol. It has no \texttt{ConsentRecord}, no \texttt{AdherenceEvent},
no versioned chain, and no capability-bound invalidation. ACAP uses ODRL
vocabulary for claim semantics and addresses the consent lifecycle layer
that ODRL does not cover.

Three contributions distinguish this work from the above: the
proof-of-adherence framing that shifts from one-time acceptance to
per-action, per-clause adherence events with reasoning trails; the
inter-agent consent chain as a versioned linked-list audit structure
\emph{between} two agents; and capability-bound consent that ties
consent validity to the agent's own configuration.

\begin{center}\rule{0.5\linewidth}{0.5pt}\end{center}

\subsection{6. Discussion}\label{discussion}

\subsubsection{6.1 Threat Model and
Limitations}\label{threat-model-and-limitations}

The \texttt{reasoning} field on \texttt{AdherenceEvent} is free-form
natural language. This is a deliberate choice for v0.1 such that the
field is immediately useful to human auditors. It also means reasoning
strings are not machine-comparable; two agents may evaluate the same
clause differently and produce non-equivalent reasoning strings with no
way to detect the discrepancy. A structured reasoning format would
enable automated auditing but would raise the implementation barrier
substantially, and we defer this.

The model assumes calling agents act in good faith when populating
\texttt{parsed\_claims} and \texttt{reasoning}. A malicious agent could
record \texttt{understood:\ true} for a prohibition and then violate it.
This is worse than the current state in that it produces false evidence,
but it also creates a clearer liability standard. An agent that
explicitly acknowledged a prohibition and violated it has acted in
demonstrable bad faith, which is legally more tractable than silent
non-compliance under a clicked-through ToS. Similarly, property S7
guarantees that a capability change triggers re-consent, but the
\texttt{caller\_capability\_hash} is self-reported and the callee has no
mechanism to independently verify the caller's capability fingerprint.
The honest comparison is not ``ACAP vs.~perfect enforcement'' but ``ACAP
vs.~the current state,'' which is no consent record at all.

A fuller adversarial threat model (adversarial \texttt{caller\_intent}
declaration that steers an LLM-backed claim parser toward a favourable
parse, \texttt{ConsentRecord} replay across agent identities, and
policy-squatting through an initial permissive version followed by a
restrictive version bump) is out of scope for the present work. We note
that analogous gaps exist in the human-consent layer that ACAP is
replacing, since a human signing a contract can also misstate intent and
a service can also substitute terms between acceptance and use. The
initial deployment context we anticipate is closed enterprise
integration in which both the caller's principal and the callee operate
under the same organisational trust boundary; the adversarial extensions
belong to a subsequent paper alongside remote-attestation integration
and signed replay defenses.

\subsubsection{6.2 Open Questions}\label{open-questions}

Three questions the community should resolve:

\begin{enumerate}
\def\labelenumi{\arabic{enumi}.}
\item
  \textbf{Self-attestation vs.~verification.}
  \texttt{ParsedClaim.understood} is currently self-attested. A
  verifiable proof that the calling agent processed the claim text would
  close the S7 gap; zero-knowledge proofs of document processing are
  theoretically possible but practically expensive.
\item
  \textbf{Grace periods.} \texttt{valid\_until:\ "on\_version\_bump"}
  blocks all skill invocation immediately on version change. For
  long-running tasks, an immediate block may be disruptive, and a grace
  period (for example, complete in-flight tasks, block new initiations)
  needs specifying.
\item
  \textbf{Capability fingerprint granularity.} The
  \texttt{caller\_capability\_hash} is defined as a SHA-256 digest of
  model identifier, tool manifest, and reasoning configuration. In
  practice, what constitutes a ``material'' capability change is
  ambiguous. A minor prompt template update may not affect policy
  reasoning, while a model version bump almost certainly does. The
  current design treats any fingerprint change as a re-consent trigger;
  a future version may introduce a capability diff to distinguish
  material from immaterial changes.
\end{enumerate}

\subsubsection{6.3 Extensions Under
Development}\label{extensions-under-development}

Several extensions to the core model are maintained in the accompanying
repository as separate proposals. These include tiered escalation via a
governance agent, asymmetric sensitivity preferences across data
category and usage dimension, structured regulatory context propagation
(for frameworks such as HIPAA, GDPR, PCI-DSS, and the EU AI Act), and a
plain-English audit projection layer. Each extension is maintained
independently with its own \texttt{README.md} and \texttt{STATUS.md} and
is not part of the core normative specification presented here.

\begin{center}\rule{0.5\linewidth}{0.5pt}\end{center}

\subsection{7. Conclusion}\label{conclusion}

The consent model we inherited from human authentication systems is a
timestamped acceptance record, and it is inadequate for autonomous agent
protocols. It proves acceptance. It cannot prove adherence. Agents are
capable of something humans are not, namely the ability to parse policy
documents, evaluate clauses at runtime, and produce reasoning trails for
every action. The three primitives proposed here
(\texttt{PolicyDocument}, \texttt{ConsentRecord}, and
\texttt{AdherenceEvent}) provide the infrastructure to make that
capability useful. The versioned linked-list audit chain gives legal
accountability to the calling agent's principal. The A2A and MCP
instantiations in §4 demonstrate that ACAP deploys against today's agent
protocol infrastructure without requiring any changes to the A2A or MCP
core specifications. Adoption is a matter of installing a middleware
library at each participating agent, not of advancing a specification
revision.

The timing matters. The EU AI Act's high-risk enforcement date of August
2, 2026 is months away. The A2A specification is under active revision
by a Linux Foundation working group, and the FINOS AI Governance
Framework has an open gap in MI-21 on consent versioning. This is an
unusually short window in which a protocol proposal can land in multiple
active governance processes simultaneously.

I built a version of this system for humans, years ago at a consumer
technology company. Linked consent chains, immutable versions, the whole
audit trail. Users spent four seconds on the acceptance screen and the
chain proved acceptance with precision but proved nothing else. Agents
do not need four seconds. They need a protocol that asks them to
actually read the document, evaluate each clause, and leave a record of
what they decided and why. That is what ACAP provides.

\begin{center}\rule{0.5\linewidth}{0.5pt}\end{center}

\subsection{References}\label{references}

{[}1{]} AI Agent Index 2025. arXiv:2602.17753, February 2026. (The
report covers the 2025 landscape; the preprint was published in February
2026.)

{[}2{]} EU AI Act. Regulation (EU) 2024/1689.
\url{https://digital-strategy.ec.europa.eu/en/policies/regulatory-framework-ai}

{[}3{]} EU AI Office. AI Act Service Desk: Frequently Asked Questions.
\url{https://ai-act-service-desk.ec.europa.eu/en/faq} (As of March 2026,
no FAQ or guidance document addresses consent mechanisms for autonomous
agent-to-agent transactions.)

{[}4{]} A2A Protocol Specification.
\url{https://a2a-protocol.org/latest/specification/}

{[}5{]} Model Context Protocol Specification.
\url{https://modelcontextprotocol.io/specification/}

{[}6{]} MCP November 2025 Changelog.
\url{https://modelcontextprotocol.io/specification/2025-11-25/changelog}

{[}7{]} Uniform Electronic Transactions Act (UETA), §14.
\url{https://www.uniformlaws.org/}

{[}8{]} Proskauer: Contract Law in the Age of Agentic AI, 2025.
\url{https://www.proskauer.com/blog/contract-law-in-the-age-of-agentic-ai-whos-really-clicking-accept}

{[}9{]} W3C ODRL Information Model 2.2.
\url{https://www.w3.org/TR/odrl-model/}

{[}10{]} RFC 7515: JSON Web Signature (JWS).
\url{https://datatracker.ietf.org/doc/html/rfc7515}

{[}11{]} AP2: Agent Payments Protocol.
\url{https://github.com/google-agentic-commerce/ap2}

{[}12{]} Ngozo, J.F. Open Agent Governance Specification (OAGS).
Sekuire, 2026.
\url{https://sekuire.ai/blog/introducing-open-agent-governance-specification}

{[}13{]} FINOS AI Governance Framework v2.0.
\url{https://air-governance-framework.finos.org/}

{[}14{]} Aylward, J. et al.~AIGA: AI Governance and Accountability
Protocol. IETF Internet-Draft, draft-aylward-aiga-1.
\url{https://datatracker.ietf.org/doc/draft-aylward-aiga-1/}

{[}15{]} McDonough, R. OpenMandate: Governing AI Agents by Authority,
Not Instruction. Law://WhatsNext, 2026.
\url{https://lawwhatsnext.substack.com/p/openmandate-governing-ai-agents-by}

{[}16{]} Mavračić, J. Policy Cards: Machine-Readable Runtime Governance
for Autonomous AI Agents. arXiv:2510.24383, 2025.

{[}17{]} Palumbo, N. et al.~PCAS: Policy Compiler for Secure Agentic
Systems. arXiv:2602.16708, 2026.

{[}18{]} Gaurav, S. et al.~Governance-as-a-Service: A Multi-Agent
Framework for AI System Compliance and Policy Enforcement.
arXiv:2508.18765, 2025.

{[}19{]} Wang, C.L. et al.~MI9: An Integrated Runtime Governance
Framework for Agentic AI. arXiv:2508.03858, 2025.

{[}20{]} IEEE P7012: Standard for Machine-Readable Personal Privacy
Terms. \url{https://standards.ieee.org/ieee/7012/}

{[}21{]} W3C Data Privacy Vocabulary (DPV) v2.2.
\url{https://w3c.github.io/dpv/dpv/}

{[}22{]} Kantara Initiative: Consent Receipt Specification v1.1.
\url{https://kantarainitiative.org/}

{[}23{]} Li, D., Yu, G., Wang, X. and Liang, B. AuditableLLM: A
Hash-Chain-Backed, Compliance-Aware Auditable Framework for Large
Language Models. Electronics, 15(1), 56. MDPI, 2025.

{[}24{]} Rida, C. When an AI Agent Says `I Agree,' Who's Consenting?
TechPolicy.Press, December 2025.
\url{https://www.techpolicy.press/when-an-ai-agent-says-i-agree-whos-consenting/}

{[}25{]} RFC 9396: OAuth 2.0 Rich Authorization Requests.
\url{https://datatracker.ietf.org/doc/html/rfc9396}

{[}26{]} Kantara Initiative. \emph{User-Managed Access (UMA) 2.0 Grant
for OAuth 2.0 Authorization}. January 2017.
\url{https://kantarainitiative.org/uma-specifications/}

\end{document}